\newcommand{\R}{\mathds{R}}

\def\ddr#1{\frac{\mathrm d#1}{\mathrm dr}}
\def\dpar#1#2{\frac{\partial #1}{\partial #2}}
\def\dtot#1#2{\frac{\mathrm d#1}{\mathrm d#2}}

\def\S{{\mathfrak S}}
\newcommand{\omcrit}{\omega_{\mathrm{crit}}}

\documentclass[12pt,oneside]{amsart}

\usepackage{amsmath}               
\usepackage{times}
\usepackage{hyperref}
\usepackage{epsfig}
\usepackage{dsfont}

\numberwithin{equation}{section}

\title[Global visibility of naked singularities]%
{Global visibility of naked singularities}

\author[R.\ Giamb\`o]{Roberto Giamb\`o}
\address{Dipartimento di Matematica e Informatica,
Universit\`a di Camerino, Italy} \email{roberto.giambo@unicam.it}
\urladdr{http://dmi.unicam.it/\~{}giambo}

\theoremstyle{plain}\newtheorem{teo}{Theorem}[section]
\theoremstyle{plain}\newtheorem{prop}[teo]{Proposition}
\theoremstyle{plain}
\theoremstyle{plain}
\theoremstyle{definition}
\theoremstyle{remark}\newtheorem{rem}[teo]{Remark}
\theoremstyle{definition}

\theoremstyle{plain}

\begin{document}

\begin{abstract}
Global visibility of naked singularities is analyzed here for a
class of spherically symmetric spacetimes, extending previous
studies - limited to inhomogeneous dust cloud collapse - to more
physical valid situations in which pressures are non-vanishing.
Existence of nonradial geodesics escaping from the singularity is
shown, and the observability of the singularity from far-away
observers is discussed.
\end{abstract}

\maketitle

\section{Introduction}\label{sec:intro}
The study of gravitational collapse of spherically symmetric solutions in
General Relativity led to many examples of locally naked singularities,
starting from pioneering works in the early 80's -- a quite exhaustive and
updated list of references can be found in reviews \cite{giannoni, joshi}. Most of these papers
concentrate on the aim to find photons emanated from the singularity and
escaping the Schwarzschildian trapped region at least locally. Moreover, the
analysis has been always limited to the study of \emph{radial} null geodesics,
quite simplifying then the system to study, which remains not defined at the singular point, but fully decouples.

However, in order to study the effects of these photons for a
distant observer, global behavior of null geodesics must be
studied in full generality, since \emph{nonradial} geodesics, with
nonzero angular momentum, determine the angular diameter of the
central naked singularity as seen by the observers, giving the
measure of the "size" of the singularity. In this direction there
are in literature quite detailed studies of Tolman--Bondi self
similar dust cloud collapse. In  \cite{djd, nakao} necessary
conditions -- which actually turn out to be sufficient -- for
nonradial geodesics existence are derived, and behavior at
infinity of these photons is numerically studied for some
particular case. Further, a complete result of nonradial geodesic
existence but under the assumption of self similarity of the dust
solution is fully proved in \cite{nolan}, and considerations about
topology of the singularity are derived.

Of course, the dust model cannot be considered as a good physical
model of singularity formation, because pressures are expected to
occur during the collapse. In this paper we extend the results on
dust models  to a wide class of solutions, found in \cite{ns}, for
which local nakedness results from a suitable choice of initial
data. This class represents the wider set found so far of locally
naked singularities in the gravitational collapse of elastic
materials. Since these models are not pressureless, and in
particular radial pressure does not generally vanish along a
timelike hypersurface, a junction between a solution from this
class and a anisotropic generalization of de Sitter space time
discussed in \cite{aniso} is performed to construct a global
model. It is found that properties satisfied by dust solution and
showed in previously cited works remains qualitatively valid for
this wider class of collapsing spacetimes.

The paper is organized as follows. Interior and exterior solutions are
described in Section \ref{sec:sol}, and local existence of nonradial geodesic -- which are continued in the
exterior spacetime using junction conditions --
is derived in Section \ref{sec:nonrad}. The resulting behavior for a
distant observer is analyzed in Section \ref{sec:far}, together with some remarks about photons with infinite
redshift. Last Section \ref{sec:final} is devoted to overall conclusions.

\section{Collapse of spherical anisotropic matter}\label{sec:sol}
In this part we briefly review the model we will deal with for the
rest of the paper.
The spacetime will be made by two different solutions to Einstein field
equations: an interior part, that collapses until singularity forms, and an
exterior part, that is matched with the former at a timelike hypersurface in
order to satisfy Israel--Darmois conditions \cite{israel}; it amounts to say that
both first and second fundamental forms induced by the two
solutions on this hypersurface respectively coincide.
The interior part is described by the collapsing
anisotropic solutions found in \cite{ns}, which will be matched
with the de Sitter generalizations discussed in \cite{aniso} (see
also references therein).

\subsection{Interior region}\label{sec:ns}
The interior region is provided by a class of anisotropic collapsing
matter. To describe it, the use of area--radius coordinates, first
introduced by Ori \cite{ori}, turns out to be more useful than usual
comoving reference frame
\begin{equation}\label{eq:dscom}
\text ds^2=-e^{2\nu}\text dt^2+\frac 1\eta\text dr^2+R^2(\text
d\theta^2 + \sin^2\theta\,\text d\varphi^2).
\end{equation}
Indeed, as well known \cite{magli}, the general spherical matter
distribution can be given assigning energy density $\epsilon$ and
pressures $p_r$ and $p_t$ as suitable functions of the triple
$(r,R,\eta)$. Under the assumption $\frac{\partial p_r}{\partial
\eta}=0$, in \cite{ns} it is shown that the metric takes the form
\begin{subequations}
\begin{equation}\label{eq:dsint}
\text ds^2=-A(r,R)\text dr^2 - 2 B(r,R) \text dR\text dr -C(r,R)
\text dR^2 +R^2 (\text d\theta^2 + \sin^2\theta\,\text d\varphi^2),
\end{equation}
where
\begin{equation}\label{eq:gij}
A=\left(1-\frac{2\Psi}R\right)G^2,\quad B=-G\, \frac Yu,\quad
C=\frac 1{u^2}.
\end{equation}
In \eqref{eq:gij}, the two functions $\Psi(r,R)$ and $Y(r,R)$ are
arbitrary (positive) functions while
\begin{equation}\label{eq:MS}
u^2=Y^2+\frac{2\Psi}R -1,
\end{equation}
and the function $G$ is given in terms of a quadrature:
\begin{equation}\label{eq:G}
G(r,R)=\int_R^r\frac{1}{Y(r,\sigma)}\dpar{(1/u)}r(r,\sigma)\,\text
d\sigma+\frac{1}{Y(r,r)u(r,r)}.
\end{equation}
\end{subequations}
In particular, the function $\Psi$ represents Misner--Sharp mass.
Conditions on Taylor developments of the above quantities may be
given to characterize complete collapse and naked singularity
formation, that we summarize in the following theorem.

\begin{teo}[\cite{ns}]\label{thm:ns}

\begin{enumerate}
\item In the spacetime described by the metric \eqref{eq:dsint}--\eqref{eq:G}, the singularity forms at the centre $r=0$ in a finite amount of comoving time if Taylor development of the function $R u^2$ is as follows:
\begin{equation}
Ru^2=\alpha r^3+\beta r^2 R+\gamma r R^2+\delta
R^3+o(r^2+R^2)^{3/2},\qquad(\alpha>0).\label{eq:H}
\end{equation}
\item Under the above condition,
and introduced the Taylor development of the (regular) function
$G(r,0)$
\begin{equation}\label{eq:G0}
G(r,0)=\xi r^{n-1}+o(r^{n-1}),
\end{equation}
then the central singularity is (locally) naked if, $n=1$, $n=2$, or
$n=3$ and $\xi> \alpha\,\omcrit$ where
$\omcrit=\frac{26+15\sqrt{3}}{2}$.
\end{enumerate}
\end{teo}

We also remark that this model possesses nonvanishing anisotropic pressures, which
are given by \cite{ns}
\begin{align}
&p_r=-\frac{1}{4\pi R^2}\dpar\Psi R,\label{eq:pr-now}\\
&p_t=-\frac{1}{8\pi
u\,R\,G}\left(\frac{1}{Y}\dpar{{}^2\Psi}{r\partial R}-\frac{1}{Y^2}\dpar\Psi r\dpar
YR+u\,G\dpar{{}^2\Psi}{R^2}\right), \label{eq:pt-now}
\end{align}
and in this sense the model can be considered physically more reasonable than
Tolman--Bondi dust collapsing sphere, where the spacetime is ruled by the
pressureless equation of state $p_r=p_t=0$.

\subsection{Exterior region}\label{subsec:deS}
In our model, since the internal source is given by
\eqref{eq:dsint}, one cannot hope, in general, that radial pressure
vanishes at some timelike hypersurface
$\Sigma=\{(t,r,\theta,\phi)\,:\,r=r_b\}$.
This is a quite restrictive feature of dust cloud collapsing model \cite{nolmena}
or also non vanishing radial pressure models \cite{magli}, that here arise
only as very special cases, that happen when $\Psi=\Psi(r)$.
Therefore, it cannot be
possible to consider Schwarzschild vacuum solution as external
region, if we want Israel--Darmois condition to be satisfied.

Hence,
we perform a junction between the internal source satisfying
conditions stated in Theorem \ref{thm:ns}, and the anisotropic
generalization of de Sitter spacetime, which are a class of
spherically symmetric solutions of Einstein equation satisfying the
condition $\epsilon+p_r=0$ and admitting a particular $G_4$ group of
motions (see \cite{aniso}). A coordinate transformation exists, that
brings the line element in the form
\begin{equation}\label{eq:KS}
\text ds^2=-\chi(R)\,\text dT^2+\chi(R)^{-1}\,\text dR^2+R^2\,\text
d\Omega^2,\quad\chi(R)=1-\frac{2M(R)}R,
\end{equation}
thereby obtaining a family of solutions as Misner--Sharp mass $M(R)$
varies. In \cite{homosf} it is shown that, actually, \emph{any}
spherically symmetric line element \eqref{eq:dscom} can be matched
with a metric of this family at a timelike hypersurface $\Sigma$ as
before, under the condition of continuity of mass only. Therefore,
it suffices to choose
\[
M(R)=\Psi(r_b,R),\qquad\forall R\in[0,r_b]
\]
and junction conditions will be certainly satisfied
for all (comoving) times $t\ge 0$. The value of the external mass for bigger
values of $R$ depends on the internal spacetime at comoving times prior to
observation starting, and we will suppose that it is chosen such that
\begin{equation}\label{eq:massinfty}
\limsup_{R\to+\infty} M(R)<+\infty.
\end{equation}

\subsection{Energy condition}\label{sec:ec}
In order to deal with a physically reasonable class of solutions,
we impose the weak energy condition (w.e.c.) on the
energy--momentum tensor $T$ of the spacetime. Basically, this
means $T(v,v)\ge 0$ for all timelike vectors $v$, and it is easily
seen \cite{ns} that it holds in the internal region if, for each
$r\in[0,r_b]$, the two functions of the variable $R$ only
\begin{equation}\label{eq:wec1}
R\mapsto\frac{1}{Y(r,R)}\dpar\Psi r(r,R),\qquad R\mapsto\dpar\Psi{R}(r,R)
\end{equation}
are nonnegative subsolutions of the \emph{same} ODE:
\begin{equation}\label{eq:wec2}
\frac{\mathrm dF}{\mathrm dR}\le \frac 2R\, F(R)
\end{equation}
for $R>0$. In particular, for $r=r_b$, the above condition
ensures w.e.c in the external region also (see \cite[eq. (4.6)]{homosf}).

Models of internal solutions satisfying \eqref{eq:H}, \eqref{eq:massinfty}
and w.e.c. can be easily found: for instance,
the choice
\begin{equation}\label{eq:TBdS}
\Psi(r,R)=\int_0^r\gamma(s) s^2\,\mathrm ds+\int_0^R\chi(\sigma)\sigma^2\,\mathrm d\sigma,\qquad Y=Y(r),
\end{equation}
with $\gamma(s)$ and $\chi(\sigma)$ positive and not increasing in $[0,+\infty)$,
allows for \eqref{eq:H} and the
weak energy condition to be satisfied. As an example, one may take
\[
\chi(\sigma)=\frac1{(1+R^3)^4}
\]
in order to satisfy also \eqref{eq:massinfty}. Note that, with the choice of
$\chi=\mathrm{constant}$,
the spacetime coincides with
the so-called Tolman–-Bondi–-de Sitter (TBdS), and that's the reason why, in
\cite{ns}, the models arising from the above choice \eqref{eq:TBdS} are termed
\emph{anisotropysations of TBdS spacetime}.

\section{Null geodesics from the singularity}\label{sec:nonrad}
\subsection{Geodesic equations}
As stated in Theorem \ref{thm:ns}, conditions on the metric
functions allow to determine when the central singularity is locally
naked. This is made by showing the existence of a future pointing
radial null geodesic which lies in the region $R>2\Psi$ and may be
traced back to the central singularity. In \cite{ns} it is
shown that, to obtain violations of cosmic censorship in spherical
symmetry, one may restrict oneself in looking for null geodesic
which are radial only. In other words, if a singularity is radially
censored, it is censored all the way. Nevertheless, if one wants to
study visibility under a more general point of view, also non radial
light rays should be taken into account.

Let $\kappa$ be the affine parameter of the null geodesic. Without
loss of generality, we will suppose that the geodesic lies in the
hypersurface $\theta=\frac\pi 2$, and then the angular components of
tangent vector along the geodesic read

\begin{equation}\label{eq:kang}
\kappa^\theta:=\frac{\mathrm d\theta}{\mathrm d\kappa}=0,\qquad
\kappa^\phi:=\frac{\mathrm d\phi}{\mathrm d\kappa}=\frac{\ell}{R^2},
\end{equation}
where $\ell$ is the (conserved) angular momentum. Let us also
introduce a function $q$ such that
\begin{equation}\label{eq:kr}
\kappa^r:=\frac{\mathrm dr}{\mathrm d\kappa}=\frac{1}{q\,R\,G}.
\end{equation}
Then, expressing $R$, $\theta$ and $\phi$ as functions of $r$
instead of $\kappa$, the equations for null geodesics can be given in the
following form:
\begin{subequations}
\begin{align}
&\frac{\mathrm dR}{\mathrm dr}=u\,G\left[Y-u\sqrt{1+\left(\frac{\ell\,q}{u}\right)^2}\right],\label{eq:Rr}\\
&\frac{\mathrm d\theta}{\mathrm dr}=0,\quad\frac{\mathrm
d\phi}{\mathrm dr}=\frac{\ell\,q\,G}{R},\label{eq:angr}
\end{align}
and
\begin{multline}\label{eq:2nd}
-\frac1q\ddr q-\frac1G\ddr G-\frac1R\ddr R-\frac{C\,A_{,r}+B(A_{,R}-2B_{,r})}{2G^2}\\
-\frac{C\,A_{,R}-B\,C_{,r}}{G^2}\ddr
R+\frac{B\,C_{,R}+C(C_{,r}-2B_{,R})}{2G^2}\left(\ddr
R\right)^2+\ell^2\frac{q^2\,B}{R}=0,
\end{multline}
\end{subequations}
and $A,B,C$ are given in \eqref{eq:gij}. Dealing with radial
geodesics results in vanishing of $\ell$, and so \eqref{eq:Rr},
which is found imposing that the geodesic is null, becomes an ODE in
the $R$ function only, which is enough to study at least the
existence of corresponding pregeodesics. On the other side, when
$\ell\ne0$, also \eqref{eq:2nd} must be studied.

Before to state and show results, we restrict the analysis hereafter
to the case when $n$ in \eqref{eq:G0} is equal to 3.
As it will be cleared in Remark \ref{rem:strength}, this particular situation
-- that, as seen in the statement of Theorem \ref{thm:ns}, has a sort of
``endstate transition" -- corresponds to a so--called \emph{strong curvature}
singularity, unlike $n=1,2$ cases -- at least along null radial geodesics.
We refer the reader to Remark \ref{rem:strength} below
and, for instance, to \cite{clarkrol,tip,tipclark} for a general insight
about strong curvature conditions.

\subsection{Null geodesics existence}
Existence of null radial geodesics has already been proved in
\cite{ns}, using comparison arguments in ODE. We here complete the
analysis about the asymptotic behavior of the geodesics finding also an
existence result for nonradial geodesics.

\begin{rem}\label{rem:ODE}
In the forthcoming Proposition it will be shown that the null geodesic
equation can be put in the form
\begin{equation}\label{eq:ODE}
\dtot yr=\frac1r f(y)+g(r,y),
\end{equation}
where $y(r)\in\R^n$ and
$f:\R^n\to\R$, $g:\R^{n+1}\to\R$ are $C^1$ functions;
Introducing a new independent variable $s$ such that
$\dtot rs=-r(s)$ then, from \eqref{eq:ODE}, we obtain the system
\begin{equation}\label{eq:system}
\begin{cases}
&\dtot ys =-f(y(s))-r(s)\,g(r(s),y(s)),\\
&\dtot rs=-r(s),
\end{cases}
\end{equation}
whose solutions describe parameterizations of solutions of \eqref{eq:ODE} by the parameter
$s$.
Equilibria of the above system (as $s\to+\infty$)
are given by points $(y_0,0)$ where $y_0$ is a root of
$f(y)=0$, and searching for these equilibria
amounts to search for admissible solutions of the so--called \emph{root equation},
first introduced in \cite{dj} for the study of radial light rays.

The key point is that these equilibria for the null geodesic equation will be
showed to be hyperbolic -- see e.g. \cite{palis} for basic concepts about hyperbolic dynamical systems --
and that is the reason why the
root equation solution's existence is a necessary but also sufficient
condition for light rays existence.

\end{rem}

\begin{prop}\label{thm:rng}
Under the hypotheses of Theorem \ref{thm:ns}, when , with reference
to equation \eqref{eq:G0}, $n=3$ and $\xi>\alpha\omcrit$, there exists
infinite null geodesics emanating from the central singularity and
escaping  from the trapped region $\{R>2\Psi\}$. In particular, there exists
three numbers $x_1$, $x_2$, $x_c$, depending on $\alpha$ and $\xi$, with $x_c>x_2>x_1>\alpha$,
such that these geodesics can be divided in the following classes:
\begin{enumerate}
\item\label{itm:one} infinite radial and non radial geodesics such that $R(r)= x_1
r^3+o(r^3)$,
\item\label{itm:two} infinite non radial null geodesic such that $R(r)= x_2
r^3+o(r^3)$, and
\item\label{itm:cauchy} a radial geodesic (Cauchy horizon) such that $R_c(r)=x_c r^3+o(r^3)$,
which bounds from above any
geodesic in classes \eqref{itm:one} and \eqref{itm:two}, in the sense that $R_c(r)>R_g(r),\forall r>0$, for any other
geodesic $R_g(r)$.
\end{enumerate}

Moreover, the nonradial geodesic of case \eqref{itm:one} have finite $\lim_{r\to
0^+}\phi(r)$, whereas nonradial geodesics of case \eqref{itm:two} are such
that $\lim_{r\to 0^+}\phi(r)=-\infty$.
\end{prop}

\begin{proof}
Let us consider system \eqref{eq:Rr}--\eqref{eq:2nd}. We are
interested in determining existence of solutions such that $R(r)$
lies above the apparent horizon $R_h(r)$ , which is known (see
\cite{ns} for details) to have the behavior $R_h(r)=\alpha
r^3+o(r^3)$. Then, we first introduce a new unknown function $z(r)$
in place of $R$, which is defined as
\begin{equation}\label{eq:x}
z(r)=\sqrt{\frac{R(r)}{\alpha r^3}},
\end{equation}
and we will study the system made by the first and the last equation
above, in the unknown functions $(z(r),q(r))$ -- with $q(r)$ given by \eqref{eq:kr} --
since the equation for $\phi$ can be decoupled from these. Moreover, note that
since $\ell$  appears in \eqref{eq:Rr}--\eqref{eq:2nd} as a factor of the quantity $\ell q$
only,
we will consider $\ell q(r)$ instead of $q(r)$ as variable.

We first observe that $z(r)$ and $q(r)$ must be bounded from below
by $1$ and 0, respectively, in order for the solution to be
physically acceptable. Then, estimates of terms involved in
\eqref{eq:Rr} will be performed, under the additional hypothesis
that $z(r)$ is also bounded from below, and equations \eqref{eq:Rr}
and \eqref{eq:2nd} become, with the further position
\[
\omega=\frac\xi\alpha,
\]
\begin{subequations}
\begin{align}
&\ddr z=\frac1r w(z,\ell q)+ g_1(r,z,\ell q),\label{eq:geo1}\\
&\ell\ddr {q}=\frac1r\left[\ell q\left(k(z)-(\ell q)^2h(z)+w(z,\ell q) \widetilde k(z)\right)\right]+
\ell q\,g_2(r,z,\ell q),\label{eq:geo2}\\
\intertext{where}
&w(z,\ell q)=-\frac{2z^4+\sqrt{1+(\ell q)^2 z^2}z^3-\omega\,z+\sqrt{1+(\ell q)^2z^2}\,\omega}{2z^3},\label{eq:w}\\
&k(z)=\frac{\omega^2-10 z^3\omega-2z^6}{2z^3(\omega+z^3)},\label{eq:k}\\
&h(z)=\frac{\omega+z^3}{z},\label{eq:h}\\
&\widetilde
k(z)=-\frac{z^3+4\omega}{z(\omega+z^3)},\label{eq:ktilde}
\end{align}
\end{subequations}
and $g_i(r,z,\ell q)$, $i=1,2$ are continuous functions.

Let us check equilibria of \eqref{eq:geo1}--\eqref{eq:geo2}, using the
idea from Remark \ref{rem:ODE}.
The following cases happen.

\begin{enumerate}
\item[(a)]\label{itm:q0} $\ell q=0,\,z$ such that $w(z,0)=0$;
\item[(b)]\label{itm:q1} $\ell q=\sqrt{\frac{k(z)}{h(z)}},\,z$
such that $w(z,\sqrt{\frac{k(z)}{h(z)}})=0$.
\end{enumerate}

First, in both cases it can be seen that $(g_1(r,z,\ell q), \ell q\, g_2(r,z,\ell q))$ is regular in a neighborhood
of $(0,z,\ell q)$ where $(z,\ell q)$ is an equilibrium as above. Let us now discuss the
character of these equilibria.

In case (a), the situation goes as follows. With
reference to the notation of Remark \ref{rem:ODE}, the Jacobian
computed in the equilibrium $(z,0)$ is given by
\begin{equation}\label{eq:J1}
J_{(z,0)}f=
\begin{pmatrix}
  \dpar wz(z,0) & \dpar wq(z,0) \\
  0 & k(z) \\
\end{pmatrix},
\end{equation}
and one finds that there exists a \emph{root function}
\begin{equation}\label{eq:root}
\rho(z)=\frac{2z^4+z^3}{z-1}
\end{equation}
such that
\begin{subequations}
\begin{align}
&w(z,0)=0\,\Leftrightarrow\,\rho(z)=\omega,\label{eq:rho}\\
&\dpar wz(z,0)=-\frac{1}{2z^2}\rho'(z).\label{eq:rhoprime}
\end{align}
\end{subequations}

Moreover, $k(z)>0$ if and only if $ \omega>\mu(z):=(5+\sqrt{27})z^3.
$ The curves $\rho(z)$ and $\mu(z)$ are depicted in Figure
\ref{fig:1}.
\begin{figure}
\begin{center}
\psfull \epsfig{file=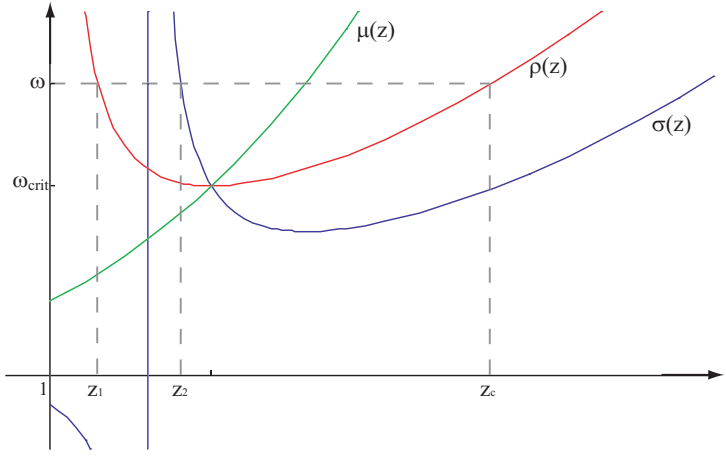, height=6cm} \caption{The curves
$\rho(z)$ \eqref{eq:root}, $\sigma(z)$ \eqref{eq:root2}, and
$\mu(z)$ intersect each other at the local minimum of $\rho(z)$,
attained for $z=\tfrac{1+\sqrt3}2,\,\rho(z)=\omcrit$ (see Theorem
\ref{thm:ns}).}\label{fig:1}
\end{center}
\end{figure}
For the acceptable range of $\omega$, that is $\omega>\omcrit$ (see
Theorem \ref{thm:ns}), there exists a unique value $z_1$ that
satisfies \eqref{eq:rho} and such that $k(z_0)>0$. From
\eqref{eq:rhoprime}, we have that also $\dpar wz(z_1,0)>0$. This
means that the Jacobian \eqref{eq:J1} has two positive eigenvalues
and therefore we have part \eqref{itm:one} of the claim,
with an infinite number of radial geodesics tending to this
equilibrium point, for every $\ell\ge 0$. Note that this situation \eqref{itm:one}
cover also radial geodesic existence, that is when $\ell=0$. Using
\eqref{eq:x}, the curves $R(r)$ behaves like $R(r)=\alpha z_1^2
r^3+o(r^3)$.

If $\omega>\omcrit$, from Figure \ref{fig:1} we see that there exists another values of
$z=z_c$ satisfying $\rho(z_c)=\omega$. In this case we have two
negative eigenvalues of \eqref{eq:J1}. With reference to the system put in the form \eqref{eq:system},
the equilibrium is again hyperbolic, but the stable manifold has dimension 1,
and therefore a single null geodesic tending to the equilibrium exists.
Actually, it can be shown that this geodesic is radial, since it can be seen
that at least a radial null geodesic tending to the equilibrium exists.
Indeed, radial geodesics are solutions of the ODE \eqref{eq:geo1} under the assumption
$\ell=0$. We can easily see that $z=z_c$ is an equilibrium of \eqref{eq:geo1},
such that there exist a unique solution tending to this equilibrium.
Since radial geodesics are solutions of
\eqref{eq:geo1}--\eqref{eq:geo2} with $\ell q\equiv 0$, we conclude that the
(unique)
solution of \eqref{eq:geo1}--\eqref{eq:geo2} must be radial.
The corresponding $x=\alpha z_c^2$ represents \emph{Cauchy horizon} direction.

We also note that the corresponding function $R_c(r)$ must bound from above
any other geodesic $R_g(r)$. Of course, this must be true
for any other radial geodesic near $r=0$, and therefore for any other value of
$r$ by uniqueness of Cauchy problem solution.
Let us show that this also holds true for nonradial geodesic. By
contradiction, let $R_g(r)$ be a nonradial geodesic, such that its projection
on the $(r,R)$ plane bounds from above $R_c(r)$ for some $r_*$, that is
$R_g(r_*)>R_c(r_*)$. Since $R_g(r)$ is a subsolution of radial null geodesic
equation, and $R_c(r)$ is a solution of the same
ODE, therefore we can find a radial null geodesic $R_n(r)$ with initial data
$R_n(r_*)\in]R_c(r_*),R_g(r_*)[$, and trace it back until $R_n(0)=0$,  obtaining a radial geodesic
$R_n(r)>R_c(r)$ in a right neighborhood of $r=0$, which as said before is
absurd, and claim \eqref{itm:cauchy} is complete.

\bigskip

Let us now discuss case (b) to get claim \eqref{itm:two} of the Proposition.

We first observe that this time it must be $k(z)>0$ otherwise we have no
equilibria. Arguing as in the previous case, we find that the
determinant of the Jacobian in the equilibrium -- say
$(z_2,\sqrt{\frac{k(z_2)}{h(z_2)}})$ -- is given by
\begin{equation}\label{eq:J2}
-2k(z_2)\dtot{w(z,\sqrt{\frac{k(z)}{h(z)}})}{z}|_{z=z_2}.
\end{equation}
The root function in this case is given by
\begin{equation}\label{eq:root2}
\sigma(z)=\frac{4z^5}{2z^2-3},
\end{equation}
which has the properties
\begin{subequations}
\begin{align}
&w(z_2,\sqrt{\frac{k(z_2)}{h(z_2)}})=0\,\Leftrightarrow\,\sigma(z_2)=\omega,\label{eq:sigma}\\
&\dpar
wz(z_2,\sqrt{\frac{k(z_2)}{h(z_2)}})=-\gamma^2\sigma'(z_2),\label{eq:sigmaprime}
\end{align}
\end{subequations}
where $\gamma$ is some non vanishing constant. The situation is
depicted in Figure \ref{fig:1}. For any $\omega>\omcrit$ there exists
a unique value of $z_2$ satisfying \eqref{eq:sigma} with $k(z_2)>0$.
Using \eqref{eq:sigmaprime}, we get that the quantity \eqref{eq:J2}
is negative and then there exists one and only one positive
eigenvalue of the Jacobian, which results in the existence of infinite non
radial null geodesic such that the corresponding $R(r)$ behaves like
$R(r)=\alpha z_2^2 r^3+o(r^3)$. Indeed, the stable manifold of the system put in the form
\eqref{eq:system} has dimension 2.

Notice that, in this case, there are no other choice
available since the other root of the equation $\sigma(z)=\omega$ lies
in the $\{k<0\}$ region.

To complete the proof,
let us now analyze the behavior of the angular function $\phi(r)$. Using
\eqref{eq:angr}, and the above estimates,
\begin{equation}\label{eq:phir}
\dtot\phi r\cong\ell \left(\frac{h(z)}z\right)\frac qr.
\end{equation}
For a non radial geodesic from case \eqref{itm:one}, $q\to 0$ as $r\to 0^+$, and then, from
\eqref{eq:geo2}, $\dtot qr\cong k(z_1)\tfrac qr$, from which
$q(r)\cong r^{k(z_1)}$, and then, using \eqref{eq:phir}, $\phi(r)$
has a finite limit as $r\to 0^+$.

For a non radial geodesic from case \eqref{itm:two} conversely, $q$ tends to a finite positive
value and then the function $\phi(r)$ goes like $k\,r^{-1}$, which
determines a negative diverging behavior of the function $\phi(r)$.
\end{proof}

\begin{rem}\label{rem:phi}
These results are completely consistent with the analysis carried
out in \cite{nakao, nolan}, which have established a double topological nature
of the naked singularity in the particular case of dust solution under
self--similarity assumption. In particular,
case \eqref{itm:one} corresponds to a region of the
singularity foliated by a 2--sphere, whereas case \eqref{itm:two}
corresponds to a topologically pointwise singular region.
\end{rem}

\begin{rem}\label{rem:strength}
If we look at the behavior of the quantity $k^2 R_{\alpha\beta}
K^\alpha K^\beta$ along null radial geodesics, where
$R_{\alpha\beta}$ is Ricci tensor, $K^\alpha=\tfrac{\mathrm
dx^\alpha}{\mathrm dk}$ is the tangent vector of the geodesic with
parameter $k$, we get
\begin{equation}\label{eq:strength}
k^2\,R_{\alpha\beta}K^\alpha K^\beta=\frac{k^2\Psi_{,r}}{4\pi R^2
u^3 G\,Y}(K^R)^2.
\end{equation}
In view of the above theorem, equations \eqref{eq:kr} and
\eqref{eq:Rr} ensure that there exists $\lim_{k\to 0^+}K^R$, then
$\tfrac{k}{R}K^R$ tends to a finite nonzero limit value. Since both $G$ and $\Psi_{,r}$ goes like
$r^2$, and $Y,u$ tends to a finite nonzero limit along the
geodesics, the quantity in \eqref{eq:strength} goes to a finite
nonzero value. It follows that these null geodesics terminate in a
strong curvature singularity in the sense of Tipler \cite{tip}.
\end{rem}

\section{Faraway visibility of the singularity}\label{sec:far}

\subsection{Size of the singularity}
In the following, we will study the behavior at infinity of null geodesics emanating from the singularity.
We have already seen that there exists an infinite number of light rays
escaping from the trapped region;
These geodesics
arrives at the boundary of the collapsing sphere of anisotropic matter with some value $R(r_b)$, and
they have to be continued by studying the behavior of null geodesics in the
external region \eqref{eq:KS} escaping from the boundary and such that
$R=R(r_b)$.

The equation for null geodesics -- that, without loss of generality, we will
suppose to lie in the subspace $\{\theta=\tfrac\pi 2\}$ -- is given by
\begin{equation}\label{eq:rg-ext}
\dtot T\kappa=\frac\omega{\chi(R)},\quad\dtot
R\kappa=\omega\sqrt{1-\frac{\chi(R)}{R^2}\left(\frac\ell\omega\right)^2},\quad
\dtot\theta\kappa=0,\quad\dtot\phi\kappa=\frac\ell{R^2},
\end{equation}
where $\kappa$ denotes as usual the affine parameter and $\ell$ and $\omega$
are constant of motion.

As explained in \cite{nakao},
since the metric \eqref{eq:KS} does not depend
on $T$ we can uniquely associate to an observer $R=R_0$ an orthonormal basis
of the tangent space to a point $(T,R_0,\theta,\phi)$ of the outer region (i.e. such that $R_0>2M(R_0)$),
which will be denoted by
$$
\left\{e_{(T)}=\chi(R_0)^{-1/2}\partial_T,\,e_{(R)}=\chi(R_0)^{1/2}\partial_R,\,
e_{(\theta)}=R_0^{-1}\partial_\theta,\,e_{(\phi)}=(R_0\sin\theta)^{-1}\partial_\phi\right\}.
$$
Then, if we consider a null  geodesic with affine parameter $\kappa$ emanating from the singularity
passing for $R=R_0$, and denote by $\kappa^\alpha=\dtot {X^\alpha}\kappa$ its tangent vector,
this observer measures an angle $\delta$
between the light ray and the radial direction equal to
\begin{equation}\label{eq:delta}
\delta=\arctan\left(\frac{g_{\mu\alpha}e_{(\phi)}^\mu \kappa^\alpha}{g_{\mu\alpha}e_{(R)}^\mu
\kappa^\alpha}\right)=\frac\ell\omega
\sqrt{\frac{\chi(R_0)}{R_0^2-(\ell/\omega)^2\chi(R_0)}}.
\end{equation}
The above quantity depends on the geodesic, and therefore the supremum made among the set $\S$ of all singular
 geodesics detected at $R=R_0$ gives a measure of the singularity
detected by the observer. One can conceive the righthand side above, fixing $R_0$, as a
function in $(\ell/\omega)$ which results to be increasing. Therefore, the
``size" of the singularity detected by the faraway observer is related to the quantity
$$b:=\sup_\S\frac\ell\omega,$$
which can be regarded \cite{nakao} as a sort of
\emph{impact parameter}.

Using this definition, it is not hard to prove an extension of the result already proved for
the particular case studied in
\cite{nakao}.

\begin{prop}\label{thm:ip}
If the boundary of the interior region $r=r_b$ is sufficiently small, then the
non radial geodesics of largest impact parameters emanating from the singularity
are such that the angular function $\phi\to-\infty$ in the approach
to the singularity (case \eqref{itm:two} of Proposition \ref{thm:rng}, see also Remark
\ref{rem:phi}).
\end{prop}

\begin{proof}
Of course, we will consider only nonradial geodesics,
when $\ell\not=0$.

Using continuity of the metric along the junction hypersurface, one gets
\begin{equation}\label{eq:basis}
\dtot T\kappa=\dtot TR\dtot R\kappa+\dtot Tr\dtot
r\kappa=\left[-\frac{Y}{\chi\,u}\dtot Rr+G\right]\dtot r\kappa,
\end{equation}
and using \eqref{eq:rg-ext}, together with \eqref{eq:kr}--\eqref{eq:angr}
one finds the angular frequency
$\omega$ and therefore the impact parameter $b$, which is given by
\begin{equation}\label{eq:b}
b=\sup_\S\left(\frac{R(r)\,\ell q(r)}{Y(r,R(r))\sqrt{u^2(r,R(r))+(\ell
q(r))^2}-u^2(r,R(r))}\right)|_{r=r_b}.
\end{equation}
With reference to the cases listed in Proposition \ref{thm:rng}, let $R_1(r)$ be a
nonradial geodesic from case \eqref{itm:one}, and let $R_2(r)$ a (necessarily non radial)
geodesic from case \eqref{itm:two}.
It is easy to verify that, as $r\to 0^+$,
the quantity in round brackets in \eqref{eq:b} tends to 0
along $R_1(r)$ and tends to a finite nonzero value along $R_2(r)$.
Then, if $r_b$ is sufficiently small, we can suppose that this quantity, along
$R_2$, bounds from above the same quantity computed along $R_1$ until
$r=r_b$.
\end{proof}

\subsection{Redshift}
Following \cite{dwi}, the frequency shift $z$ between a source and an observer respectively located at events $P_1$
and $P_2$, connected by a null geodesic with tangent vector $\overrightarrow{\kappa}$ w.r.t. the
affine parameter, is defined as
\begin{equation}\label{eq:redshift}
1+z=\frac{g_{P_1}(\overrightarrow{\kappa},u_{(s)})}{g_{P_2}(\overrightarrow{\kappa},u_{(o)})}.
\end{equation}
where $u_{(s)}$ and $u_{(o)}$ are the 4--velocities of the source and the
observer respectively.

To determine the redshift associated to the singular geodesics, the numerator
above will be replaced by a limit expression as $P_1$ approaches the central
singularity. Let us therefore consider a singular geodesic passing at event $P_2$
in the exterior space such that $R(P_2)=R_0$. Using the basis
\eqref{eq:basis}, together with \eqref{eq:rg-ext}, it is easy to see that the
denominator in \eqref{eq:redshift} is given by $-\chi(R_0)^{1/2}\omega$, which
under the hypothesis \eqref{eq:massinfty} is always finite also if $R_0$ tends
to infinity. On the other side, the velocity field in the interior spacetime
is given by $-u(r,R)\partial_R$, and using \eqref{eq:kr}--\eqref{eq:Rr} one
finds that the absolute value of the numerator in \eqref{eq:redshift} is given by
\begin{equation}\label{eq:red}
\lim_{r\to
0^+}\frac1{R(r)}\left[\left(\frac{u(r,R(r))}{q(r)}\right)^2+\ell^2\right]^{1/2}=+\infty,
\end{equation}
which results in an infinite frequency shift. Actually we can
observe that calculation of \eqref{eq:red} above seems a feature
of all singular spherically symmetric models, at least for
nonradial geodesics, when $\ell\not=0$. In the case under our
study, moreover, \eqref{eq:red} holds also for \emph{radial}
geodesic since, using Proposition \ref{thm:rng} and equation
\eqref{eq:MS}, $u(r,R(r))\cong u_0\ne 0$, and both $R(r)$ and
$q(r)$ are infinitesimal. It appears also to be possible the study
of the limit in \eqref{eq:red} in a even more general framework,
and work in this direction is in progress.

All in all, these models shows the same feature described in
\cite{dwi} for dust solutions: in the case corresponding to a
strong naked singularity (along radial geodesic\footnote{As
pointed out in \cite{dwi}, it can be possible that the strength of
the singularity may have a directional behavior, that is the
singularity is weak -- in the sense explained in Remark
\ref{rem:strength} -- along some geodesics, and strong among
others.}), null geodesics are infinitely redshifted. Therefore, it
can be said that there exists a form of weak censorship for such
class of solutions, at least at the classical level. Of course
however, quantum phenomena might change this scenario, as
suggested by some authors (see \cite{nakao},\cite{nolmena}).

\section{Discussion and conclusions}\label{sec:final}
In this paper we have studied visibility of the naked singularity
for a faraway observer, investigating global behavior of null
(radial and nonradial) geodesics in a class of spherically
symmetric singular spacetimes. The interior part belongs to a wide
class of anisotropic solutions; as known from the analysis carried
out in \cite{ns}, the endstate of the central singularity is
determined by the lower order term  $n$ of Taylor development of a
quantity determined by the metric. In this paper we considered the
case $n=3$, which has a special interest since (i) the endstate is
determined by the value of a certain parameter $\xi/\alpha$, that
causes a sort of "phase transition" between naked singularity and
black hole situations, and (ii) radial geodesics terminates into a
strong curvature singularity, as observed in Remark
\ref{rem:strength}.

The exterior part belongs to the class of anisotropic
generalization of de Sitter solution discussed in \cite{aniso}.
The form of the metric closely resembles the Schwarzschild line
element, but the mass here depends on the areal coordinate $R$
since a constant mass would not properly match an interior
solution with non vanishing radial pressures. Indeed, if
Misner--Sharp masses for both solutions (interior and exterior)
coincide at a hypersurface $r=\text{const.}$, this ensures Darmois
junction conditions to hold along this hypersurface.

Existence of nonradial geodesics is crucial to investigate global
visibility. The system of ODE satisfied by the nonradial geodesics
does not decouple and the most suitable way seems to study
equilibrium points of this system. It turns out that there are two
values $x_1<x_2$, depending on the data of the spacetime, such
that there is an infinite number of nonradial geodesic $R(r)$
emanating with "direction" $x_1$ (i.e. such that $\lim_{r\to
0}R(r)/r^3=x_1$) and infinite nonradial geodesics with direction
$x_2$. This result is not in contrast with previous results
regarding self similar collapse \cite{nolan} where one has
infinite geodesic with direction $x_1$ vs. only one geodesic with
direction $x_2$. Indeed, as the proof of Proposition \ref{thm:rng}
points out, the stable manifold of the equilibrium  has dimension
3 in the first case and 2 in the second one. Self--similarity
assumption simplifies the geometry of these stable manifolds,
shrinking them by one dimension.

It must be remarked that a crucial fact is that ODE system can be
brought in the form \eqref{eq:ODE}, where the equilibria are all
hyperbolic. This is a feature of all collapsing matter models
known in literature, which excluded pathological situations such
as the ones outlined in \cite{nolmena}, and explains why the
so-called \emph{root equation} method repeatedly exploited in
previous works -- based on the existence of a certain limit which
in principle cannot be taken for granted (see for instance
\cite{djd, dj, nakao}) -- leads all the same to the conclusions
one would expect.

The analysis of the ODE system allowed us to determine a sort of
"impact parameter" that characterizes the "size" of the
singularity as it is seen by a faraway observer. It turns out that
photons emanated by the naked singularity are infinitely
redshifted also when pressures are present, as has been already
shown (see \cite{dwi}) for the -- pressureless -- Tolman--Bondi collapse.

\end{document}